\documentclass[11pt]{article}
\usepackage{moriond,epsfig}

\def\Journal#1#2#3#4{{#1} {\bf #2}, #3 (#4)}

\def\apj{{\em Astrophys.\ J.}}
\def\apjl{{\em Astrophys.\ J.\ Lett.}}
\def\mnras{{\em MNRAS}} 
\def\nat{{\em Nature}}
\def\aap{{\em Astro.\ \& Astrophys.}}
\def\araa{{\em An.\ Rev.\ Astro.\ \& Astrophys.}}

\def\be{\begin{equation}}
\def\ee{\end{equation}}
\def\bea{\begin{eqnarray}}
\def\eea{\end{eqnarray}}

\newcommand{\G}{\Gamma}
\newcommand{\g}{\gamma}

\begin{document}
\vspace*{4cm}
\title{THE GAMMA-RAY VIEW OF THE EXTRAGALACTIC BACKGROUND LIGHT}

\author{ J.\ D.\ Finke$^{1,2}$ for the {\em Fermi}-LAT Collaboration}

\address{$^1$U.S. Naval Research Laboratory, Space Science Division, Code 7653, 
4555 Overlook Ave.\ SW, Washington, DC, 20375 USA \\  $^2$NRL/NRC Research Associate }

\maketitle\abstracts{ The Extragalactic Background Light (EBL) from
 the infrared (IR) through the ultraviolet (UV) is dominated by
 emission from stars, either directly or through absorption and
 reradiation by dust.  It can thus give information on the star
 formation history of the universe.  However, it is difficult to
 measure directly due to foreground radiation fields from the Galaxy
 and solar system.  Gamma-rays from extragalactic sources at
 cosmological distances (blazars and gamma-ray bursts) interact with
 EBL photons creating electron-positron pairs, absorbing the
 gamma-rays.  Given the intrinsic gamma-ray spectrum of a source and
 its redshift, the EBL can in principle be measured.  However, the
 intrinsic gamma-ray spectra of blazars and GRBs can vary considerably
 from source to source and the from the same source over short
 timescales.  A maximum intrinsic spectrum can be assumed from
 theoretical grounds, to give upper limits on the EBL absorption from
 blazars at low redshift with very high energy (VHE) gamma-ray
 observations with ground-based Atmospheric Cherenkov telescopes.  The
 Fermi-LAT observations of blazars and GRBs can probe EBL absorption
 at higher redshifts.  The lower energy portion of the LAT spectrum of
 these sources is unattenuated by the EBL, so that extrapolating this
 to higher energies can give the maximum intrinsic spectrum for a
 source.  Comparing this to the observed higher energy LAT spectrum
 will then give upper limits on the EBL absorption.  For blazars which
 have been detected by both the Fermi-LAT and at higher energies by
 Cherenkov telescopes, combined LAT-VHE observations can put more
 stringent constraints on the low redshift EBL.  These procedures
 above can also be reversed: for sources with an unknown redshift, a
 given EBL model and gamma-ray spectrum can lead to an upper limit on
 the source's redshift.  }

\section{Introduction}
\label{intro}

\begin{figure}
\begin{center}
\includegraphics[scale=0.34]{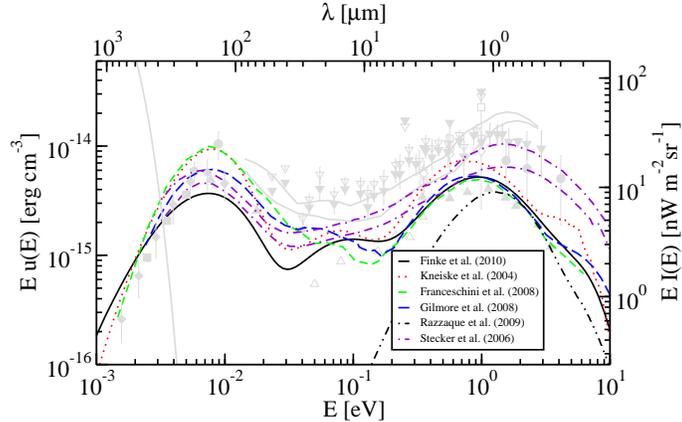}
\caption{EBL models, measurements, and constraints.  See Finke {\em et
al.} for details and references.
\label{EBLmodels_fig}}
\end{center}
\end{figure}

The night sky appears dark to the naked eye, but in fact glows faintly
in the IR through the optical and UV.  At these wavelengths, the
background light is dominated by emission from the atmosphere, solar
system, and Milky Way.  There is also a much smaller extragalactic
component from all of the stars which have ever existed, through
direct emission (in the UV-optical) and through absorption and
reradiation by dust (in the IR).  Due to the weakness of this
extragalactic background light (EBL) to other components, direct
measurement of the EBL is extremely
difficult~\cite{bernstein02,mattila03,bernstein07}.  The other
background components can be avoided to some extent by using
instruments on spacecraft which have left the
atmosphere~\cite{bernstein02,hauser98} or the solar
system~\cite{toller83,edelstein00}.  However, it is unlikely that
spacecraft will leave our Galaxy in the near future, so uncertainties
in direct measurements will remain.  Number counts in the IR and
optical can be used to find EBL lower limits~\cite{madau00,marsden09},
as discussed by Beelen and Penin in these proceedings.
Modeling~\cite{kneiske04,stecker06,franceschini08,gilmore09,finke10}
has been an important tool for constraining the EBL intensity and
tying it to basic astrophysics such as the star formation rate
density, dust absorption, initial mass function, cosmological
expansion rate, and others.  Fig.\ \ref{EBLmodels_fig} shows many EBL
measurements, constraints and models, and Hauser \&
Dwek~\cite{hauser01} present a thorough review.


The EBL photons interact with $\g$-rays from cosmological sources to
produce $e^+e^-$ pairs, absorbing the $\g$-rays so that the observed
flux $F_{obs}(E) = F_{int}(E)\exp[-\tau_{\g\g}(E)]$ where $F_{int}(E)$
is the unabsorbed source flux as a function of observed energy $E$,
and $\tau_{\g\g}(E)$ is the EBL absorption optical depth.  If
$F_{int}(E)$ is known, a measurement of the observed $\g$-ray spectrum
from these sources can be used to probe the EBL.  The intrinsic
spectrum is not generally known, however it is possible to determine
an upper limit either from theory or from extrapolating a lower
energy, unattenuated spectrum to higher energies.  This is discussed
further in the next sections.  From the upper limit on $F_{int}(E)$
and the measurement of $F_{obs}(E)$ with a $\g$-ray telescope, an
upper limit on $\tau_{\g\g}(E)$ can be calculated and compared to
theoretical predictions.

\section{Constraints with Atmospheric Cherenkov Telescopes}
\label{ACT}

Nearby blazars---active galactic nuclei with relativistic jets pointed
along our line of sight---are $\g$-ray-emitting sources up to VHE
energies and are located at cosmological distances.  They are thus a
good candidate for constraining the EBL by measuring their $\g$-ray
attenuation.  Atmospheric Cherenkov telescopes (ACTs) such as HESS,
MAGIC, and VERITAS detect $\g$-rays through the Cherenkov radiation
from particle cascades produced by $\g$-rays interacting with the
Earth's atmosphere.  TeV blazars are located nearby and VHE $\g$-rays
are generally attenuated by the mid-IR EBL.  Although they seem to be
persistent sources, they are highly variable and the intrinsic
spectrum cannot be determined.  However, theory allows the
determination of a maximum possible intrinsic spectrum.  Assuming the
$\g$-rays are produced by Compton scattering off of electrons
accelerated by na\"ive test particle acceleration theory, the hardest
possible photon index will be $\G_{int,max}=1.5$ where the photon flux
is $dN/dE\propto E^{-\G}$.  Using this, results from several blazars
(e.g. 1ES 1011-232~\cite{aharonian06}, 1ES
0229+200~\cite{aharonian07}, 3C~279~\cite{albert08}) have ruled out
high levels of the IR EBL.  However, physical mechanisms have been
suggested to produce intrinsic VHE $\g$-ray spectra harder than
$\G=1.5$~\cite{stecker07,boett08,aharonian08}.  Without a strong
constraint on $F_{int}(E)$, the constraining upper limits on the EBL
intensity are not well-accepted by some in the community.

\section{Constraints with the {\em Fermi}-LAT}
\label{LAT}

Higher $z$ sources can be probed in the GeV range using the {\em
Fermi} telescope.  The {\em Fermi} Gamma-Ray Space Telescope's primary
instrument, the Large Area Telescope (LAT) is a pair conversion
detector which surveys the entire sky every three hours in the 20 MeV
to 300 GeV range~\cite{atwood09}.  Sources located at higher $z$ will
have their VHE $\g$-ray spectrum completely absorbed; however, in the
range accessible by the LAT, these sources will have their spectra
attenuated by optical-UV EBL photons, yet not so attenuated that they
cannot be observed.  Approximately 600 blazars are listed in the first
year LAT AGN catalog~\cite{abdo10_1lac}.  Unfortunately, many of these
exhibit intrinsic spectral breaks and do not have many photons
$\geq10$ GeV needed to probe the EBL~\cite{abdo10_spectra}.  Using
statistics of the LAT blazars, as suggested by Chen et
al.~\cite{chen04} is thus not possible~\cite{abdo10_EBLconstrain}.
However, a smaller sample of blazars do not exhibit spectral breaks
and do have high energy photons, and these can be used to probe the
EBL on a case-by-case basis.  Additionally, 6 Gamma-Ray Bursts
(GRBs)\footnote{Brief, beamed $\g$-ray emission from exploding stars.}
with measured redshifts have been detected by the LAT as of this
writing (2010 May 13).  Together, these blazars and GRBs can be used
to constrain EBL models.

Abdo {\em et al.}\cite{abdo10_EBLconstrain} use two methods to do
this: the highest energy photon (HEP) method and the Likelihood Ratio
Test (LRT) method.  Both of these techniques make use of the fact that
below 10 GeV, all EBL models predict essentially no attenuation.  This
lower energy spectrum can be extrapolated to higher energies to give
the maximum possible intrinsic spectrum, which can then be compared to
the observed LAT spectrum at $>10$ GeV.  The HEP method uses a Monte
Carlo simulation to randomly draw the highest energy photon from a
distribution created with the extrapolated 0.1--10 GeV spectrum and a
particular EBL model.  Repeating this many times builds up a
distribution of HEPs which can be compared to the actual HEP observed
from a source to give the probability of rejecting the particular EBL
model used.  The LRT technique assumes as a null-hypothesis the
extrapolated 0.1--10 GeV spectrum and a certain EBL model.  A fit is
then performed with the normalization of this EBL model's opacity as a
free parameter.  From the likelihood ratio of these two fits, Wilks'
theorem can be used to determine the probability of rejecting the null
hypothesis, which is essentially the probability of rejecting the EBL
model being tested.

Combining the results from several sources, the Stecker {\em et
al.}~\cite{stecker06} baseline model is rejected with a high
significance.  It should be noted however that this rejection is only
applicable to the UV; at the mid-IR and longer wavelengths, the model
predictions are still viable.  The Stecker {\em et al.}\ fast
evolution model predicts a greater opacity than their baseline model,
and is rejected at an even greater significance.  All other EBL models
tested by Abdo {\em et al.}~\cite{abdo10_EBLconstrain} are allowed.

An additional way of using the LAT to constrain the EBL has been
suggested: using the Compton-scattering of the EBL in the lobes of
radio galaxies~\cite{georgan08}.  The recent detection of the giant
lobes of the nearby radio galaxy Cen~A by the LAT~\cite{abdo10_lobes}
demonstrates the feasability of this method, since Compton-scattering
of the cosmic microwave background alone cannot explain this emission.
Georganopoulos {\em et al.}~\cite{georgan08} suggest the
well-constrained lobes of Fornax A would be an ideal candidate for
this technique.

\section{Constraints Combining ACTs and the LAT}
\label{ACTandLAT}

The problems with theoretical uncertainties in the intrinsic VHE
$\g$-ray spectra of blazars (\S\ \ref{ACT}) can be sidestepped by
using the {\em Fermi}-LAT.  The LAT spectrum can be extrapolated into
the TeV range, giving the maximum possible TeV spectrum, assuming that
the $\g$-ray spectrum of a blazar would be concave upwards.  This
technique has been used by Georganopoulos {\em et
al.}~\cite{georgan10} with the blazar 1ES~1218+304 to reject the
Stecker {\em et al.}~\cite{stecker06} baseline and fast evolution
models with $2.6\sigma$ and $4.7\sigma$ significance, respectively,
and the ``best fit'' models of Kneiske {\em et al.}\cite{kneiske04}
with a $2.9\sigma$ significance.  Future applications of this
technique to combined LAT-ACT observations could give even stronger
constraints.

The opposite of this technique can also be used to constrain the
redshift of a blazar for which it is unknown.  Assuming a certain
model for the EBL, the VHE spectrum of a source can be deabsorbed
until it is at a higher level than the extrapolated LAT spectrum.
This technique has been used to constrain $z<0.75$ for PG
1553+113~\cite{abdo10_1553} and $z<0.66$ for PKS
1424+240~\cite{acciari10_1424}.

\section*{Acknowledgments}

The {\em Fermi} LAT Collaboration acknowledges generous ongoing support from
a number of agencies and institutes that have supported both the
development and the operation of the LAT as well as scientific data
analysis.  These include the National Aeronautics and Space
Administration and the Department of Energy in the United States, the
Commissariat \`a l'Energie Atomique and the Centre National de la
Recherche Scientifique / Institut National de Physique Nucl\'eaire et de
Physique des Particules in France, the Agenzia Spaziale Italiana and the
Istituto Nazionale di Fisica Nucleare in Italy, the Ministry of
Education, Culture, Sports, Science and Technology (MEXT), High Energy
Accelerator Research Organization (KEK) and Japan Aerospace Exploration
Agency (JAXA) in Japan, and the K.A.~Wallenberg Foundation, the Swedish
Research Council and the Swedish National Space Board in Sweden.


\end{document}